\documentclass[aps,prl,twocolumn,superscriptaddress]{revtex4}
\usepackage{graphicx}

\begin{document}


\title{Structure of $^{12}$Be: intruder $d$-wave strength at N=8}

\author{S.D. Pain}
\altaffiliation{present address: Rutgers University, c/o Physics
Division, Oak Ridge National Laboratory, TN 37831-6354.}
\affiliation{Department of Physics, University of Surrey,
Guildford, GU2 7XH, UK}
\author{W.N. Catford}
\affiliation{Department of Physics, University of Surrey, Guildford, GU2 7XH, UK}
\affiliation{Laboratoire de Physique Corpusculaire, ENSICAEN et Universit\'{e}
de Caen, IN2P3-CNRS, 14050 Caen Cedex, France}
\author{N.A. Orr}
\affiliation{Laboratoire de Physique Corpusculaire, ENSICAEN et Universit\'{e}
de Caen, IN2P3-CNRS, 14050 Caen Cedex, France}
\author{J.C. Ang\'elique}
\affiliation{Laboratoire de Physique Corpusculaire, ENSICAEN et Universit\'{e}
de Caen, IN2P3-CNRS, 14050 Caen Cedex, France}
\author{N.I. Ashwood}
\affiliation{School of Physics and Astronomy, University of Birmingham, Edgbaston,
Birmingham, B15 2TT, UK}
\author{V. Bouchat}
\affiliation{Universit\'{e} Libre de Bruxelles, CP226, B-1050 Bruxelles, Belgium}
\author{N.M. Clarke}
\affiliation{School of Physics and Astronomy, University of Birmingham, Edgbaston,
Birmingham, B15 2TT, UK}
\author{N. Curtis}
\affiliation{School of Physics and Astronomy, University of Birmingham, Edgbaston,
Birmingham, B15 2TT, UK}
\author{M. Freer}
\affiliation{School of Physics and Astronomy, University of Birmingham, Edgbaston,
Birmingham, B15 2TT, UK}
\author{B.R. Fulton}
\affiliation{Department of Physics, University of York, Heslington, York, YO10 5DD, UK}
\author{F. Hanappe}
\affiliation{Universit\'{e} Libre de Bruxelles, CP226, B-1050 Bruxelles, Belgium}
\author{M. Labiche}
\affiliation{Electronic Engineering and Physics, University of
Paisley, High Street, Paisley, Scotland, PA1 2BE, UK}
\author{J.L. Lecouey}
\affiliation{Laboratoire de Physique Corpusculaire, ENSICAEN et Universit\'{e} de
Caen, IN2P3-CNRS, 14050 Caen Cedex, France}
\author{R.C. Lemmon}
\affiliation{CCLRC Daresbury Laboratory, Daresbury, Warrington, Cheshire, WA4 4AD, UK}
\author{D. Mahboub}
\affiliation{Department of Physics, University of Surrey, Guildford, GU2 7XH, UK}
\author{A. Ninane}
\affiliation{Insitut de Physique Nucl\'{e}are, Universit\'{e} Catholique de Louvain, Louvain-la-Neuve,
Belgium}
\author{G. Normand}
\affiliation{Laboratoire de Physique Corpusculaire, ENSICAEN et Universit\'{e} de Caen,
IN2P3-CNRS, 14050 Caen Cedex, France}
\author{N. Soi\'c}
\altaffiliation{present address: Rudjer Bo\v{s}kovi\'{c} Institute, Bijeni\v{c}ka 54, HR-10000 Zagreb,  Croatia}
\affiliation{School of Physics and Astronomy, University of Birmingham, Edgbaston,
Birmingham, B15 2TT, UK}
\author{L. Stuttge}
\affiliation{Institut de Recherche Subatomique, IN2P3-CNRS/Universit\'{e} de Louis
Pasteur, BP28, 67037 Strasbourg Cedex, France}
\author{C.N. Timis}
\affiliation{Department of Physics, University of Surrey, Guildford, GU2 7XH, UK}
\author{J.A. Tostevin}
\affiliation{Department of Physics, University of Surrey, Guildford, GU2 7XH, UK}
\author{J.S. Winfield}
\affiliation{Department of Physics, University of Surrey, Guildford, GU2 7XH, UK}
\affiliation{Laboratoire de Physique Corpusculaire, ENSICAEN et Universit\'{e} de Caen,
IN2P3-CNRS, 14050 Caen Cedex, France}
\author{V. Ziman}
\affiliation{School of Physics and Astronomy, University of Birmingham, Edgbaston,
Birmingham, B15 2TT, UK}

\date{\today}

\begin{abstract}
The breaking of the $N$=8 shell-model magic number in the $^{12}$Be ground state has
been determined to include significant occupancy of the intruder
$d$-wave orbital. This is in marked contrast with all other $N$=8
isotones, both more and less exotic than $^{12}$Be. The
occupancies of the $0\hbar \omega$ $\nu p_{1/2}$-orbital and the
$1\hbar \omega$,  $\nu d_{5/2}$ intruder orbital were deduced from
a measurement of neutron removal from a high-energy $^{12}$Be beam
leading to bound and unbound states in $^{11}$Be.
\end{abstract}


\maketitle

One of the principal aims of present day nuclear structure research is to
understand the evolution of shell structure with increasing
asymmetry in the neutron-to-proton ratio. In this context the
$N$=8 isotonic chain, which spans from $^{22}$Si via the doubly
magic $N$=$Z$ $^{16}$O to the two-neutron halo system $^{11}$Li
and the unbound $^{10}$He, is of considerable interest. Indeed, the $N$=8 shell closure, that
is clearly evident close to stability, disappears amongst the lightest of these nuclei. In particular,
the halo structure of $^{11}$Li is enhanced by a strong
$\nu(1s_{1/2})^2$ intruder valence neutron configuration
\cite{HaikSimon}. Similarly, recent experiments
\cite{Iwasaki1,Navin,Iwasaki2,Shimoura1} have confirmed earlier work \cite{Barker1,Barker2,Alburger2,Fortune} in which it was
concluded that the $^{12}$Be ground state is formed from both the
``normal'' closed shell $\nu(0p_{1/2})^2$ valence configuration
and the intruder $\nu(1s0d)^2$ configurations. The factors
producing these intruder configurations appear to include a
reduction in the $p-sd$ shell gap as the dripline is approached,
an increase in the monopole pairing energy and deformation
\cite{BABProgReview}. The deformation is also believed to be related to the
tendency towards alpha-particle clustering
\cite{Kanada2}
in the Be isotopes.

Whereas high-energy single-neutron removal (or ``knockout'') from
$^{12}$Be has provided direct evidence for configuration mixing
involving the $\nu(1s_{1/2})^2$ and $\nu(0p_{1/2})^2$  valence
neutron configurations in the ground-state \cite{Navin}, model
predictions indicate that a substantial $\nu(0d_{5/2})^2$
admixture ($\sim$30-50\%) should also be present
\cite{BABProgReview,Gori,Sagawa, Nunes, VinhMau}. In the experiment of Navin {\em et al.}, $^{11}$Be
core fragments in either the $J^\pi = {1/2}^+$ ground state or
the bound 320 keV ${1/2}^-$ excited state were detected
\cite{Navin}. These states have large overlaps with the pure
single-particle states $\nu(1s_{1/2})$ and $\nu(0p_{1/2})$
respectively \cite{Zwieglinski,Fukuda1}. The measurement of
ref. \cite{Navin} was, however, not sensitive to the
$\nu(0d_{5/2})^2$ component as the removal of a
$0d_{5/2}$ neutron leaves $^{11}$Be in the $\nu(0d_{5/2})$
single-particle state (E$_{\rm x}$ = 1.78 MeV, $\Gamma$ = 100
keV) \cite{Zwieglinski,Ajzenberg1990}, which is unbound to neutron
emission and decays to $^{10}$Be(g.s.)+n.
It is thus necessary to design an
experiment to detect both the $^{10}$Be fragment and the neutron
and then to reconstruct their relative decay energy from the
measured momenta. This was the approach adopted in the present
work. Further, to assist comparisons with theory and the earlier
work, the ability to detect $^{11}$Be in the first
excited state (via the 320 keV gamma-ray) was included.

A secondary beam of $^{12}$Be ($\sim$5000 pps) was prepared using the
LISE3 spectrometer at GANIL and the reaction of a
63~MeV/nucleon $^{18}$O beam on $^9$Be. The average $^{12}$Be beam energy was 39.3 MeV/nucleon at the centre of the carbon secondary
reaction target (183 mg/cm$^2$).  The beam purity was 95\% with
the remaining 5\% being $^6$He and $^{15}$B. Owing to the poor
emittance of the secondary beam, the spot size on target was
$\sim$10mm diameter and the incident ions were tracked event-by-event using
two position-sensitive drift chambers located upstream of the carbon
target. The point of impact was thus determined to within $\lesssim 1$ 
mm at the target. The measured time-of-flight through the LISE
spectrometer allowed the $^{12}$Be ions to be selected uniquely
event-by-event from the rest of the beam particles and also provided a measure of
the energy with a resolution of 1.6\% (FWHM).

All charged particles emerging from the carbon target close to
zero degrees, including the unreacted beam, were recorded in a
telescope subtending $\pm$9$^\circ$ in the horizontal and vertical
planes. This was composed of two $50 \times 50$~mm$^2$ 500~$\mu$m
thick silicon strip detectors to measure the energy loss ($\Delta
E_{1,2}$), followed by a $4 \times 4$ array of 16 CsI stopping
detectors ($E$) 25 mm thick \cite{Ahmed}. The telescope array was
calibrated using ``cocktail'' beams containing all relevant
isotopes of Be, with several spectrometer settings to span the
energies of interest. Using $\Delta E_{1,2}$ and $E$, all observed
isotopes of H, He, Li, Be and B were clearly resolved. The silicon
detectors also provided vertical and horizontal position measurements,
which were combined with the drift chamber data to determine the
scattering angle with a resolution 0.7$^\circ$ (FWHM).

Coincident $\gamma$-rays were recorded using 4 NaI detectors
mounted around the target at angles of $\pm 45^\circ$ and $\pm
110^\circ$ to the beam, with a total absolute
photopeak efficiency of 3.5\% for detection of Doppler-shifted 320 keV $\gamma$-rays.

Neutrons were detected using the DEMON array of 91 liquid
scintillator modules \cite{Tilquin} located between
2.4 m and 6.3 m downstream of the carbon target and spanning angles out to $32^\circ$ \cite{LabichePRL}. The neutrons were distinguished
from $\gamma$-rays using standard pulse-shape discrimination and
their energies ($E_n$) were derived via the time-of-flight with a
resolution $\sim$5\%. Neutrons with $E_n \leq$ 15 MeV, originating from the target, were excluded in the analysis.

In addition to the measurements with the carbon target, data were
also acquired with no target. This determined the
background arising from $^{12}$Be beam particles that passed
through the target and reacted in the telescope. Such events gave
a degraded energy signal and could be misidentified as $^{11}$Be
or $^{10}$Be. As in previous experiments \cite{LabichePRL}, the
target-out measurements were made with the beam energy
lowered to account for the average energy loss in the target.  The background from reactions in the telescope precluded an accurate
measurement of the yield to the $^{11}$Be ground state
\cite{thesis}. For the excited states, however, coincident
detection of a $\gamma$-ray or neutron reduced substantially the
background (down to 50 and 60 \% of the target-in data
for the bound and unbound states respectively). 

\begin{figure}
\hspace{-0.5cm} \includegraphics[scale=0.35,angle=0]{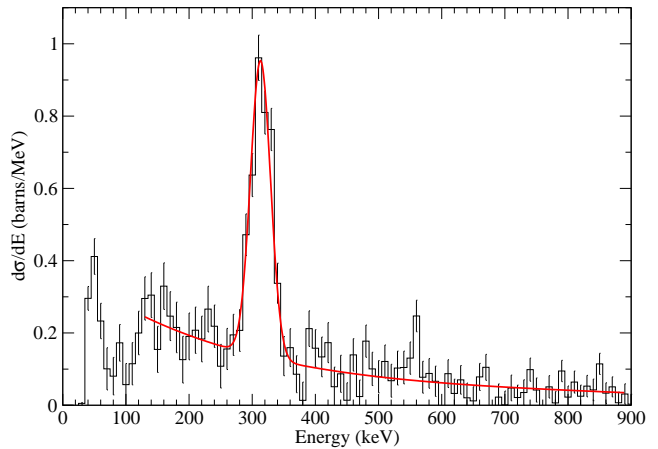}
\caption{\label{fig-Egamma} Background subtracted, Doppler
corrected $\gamma$ ray energy spectrum, in coincidence with
$^{11}$Be fragments following the reaction of $^{12}$Be (39.3
MeV/nucleon) on a carbon target. The full line is the result of a
Gaussian fit, with an exponential background. From this, the
cross section shown in Table \ref{tab-cross} was deduced for $^{11}$Be in the
first excited (${1/2}^-$) state.}
\end{figure}

The background subtracted, Doppler corrected $\gamma$-ray
spectrum, measured in coincidence with $^{11}$Be ions in the
telescope, is shown in Figure \ref{fig-Egamma}. The cross section
for production of the ${1/2}^-$ state in $^{11}$Be (see Table
\ref{tab-cross}) was extracted after taking into account the
experimentally measured $\gamma$-ray detector efficiencies,
attenuation in the target and the relativistic focussing
of $\gamma$-rays in the laboratory frame  ($\beta \simeq 0.28 c$).
The cross section measured here at 39.3~MeV/nucleon is compatible with
the value measured previously at 78 MeV/nucleon with a Be target
\cite{Navin}, as interpreted below using an eikonal reaction model. The longitudinal momentum distribution of the
$^{11}$Be*(${1/2}^-$) fragments was also measured, giving a FWHM
of 137(21) MeV/c, in agreement with the value $\sim$150 MeV/c estimated from ref. \cite{Navin}.

\begin{figure}
\hspace{0.0cm} \includegraphics[scale=0.4,angle=0]{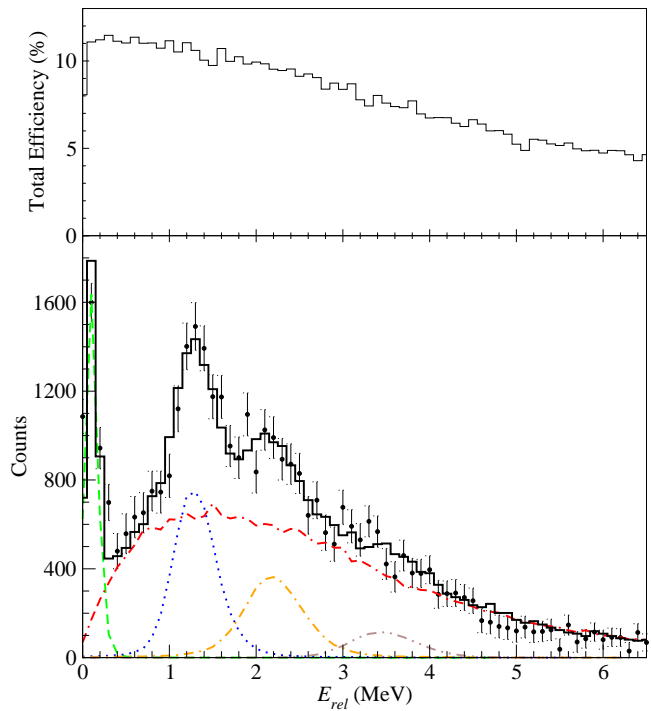}%
\vspace{0.0cm} \caption{\label{fig-Erel} 
(color online). Relative energy spectrum
of $^{10}$Be + n (lower panel) where the solid points represent the
experimental data. The histogram depicts the result of the full
simulation; the line-shapes of the individual components of the simulation
are shown (see text). The 1.78 MeV
state in $^{11}$Be is clearly visible at $E_{rel} \approx 1.3$
MeV ($S_{\textrm{n}} = 0.50$ MeV). The resolution in $E_{rel}$ varies as $a_0E^{1/2}$ (where $a_0$ is a constant), and at 1 MeV is $\sim$400 keV (FWHM). The upper panel depicts the simulated array efficiency.}
\end{figure}

Kinematic reconstruction of unbound states in $^{11}$Be was performed
from the measured momenta of coincident $^{10}$Be ions and neutrons.
The procedure was verified \cite{thesis} by reconstructing the well
known ground state resonance of $^7$He
from $^6$He + $n$ coincidences. The relative energy ($E_{rel}$)
spectrum for $^{10}$Be + n, after background subtraction, is shown
in Figure \ref{fig-Erel}. This has been corrected for the intrinsic
efficiency of the DEMON detectors but not for the geometrical
acceptance. A peak is clearly evident at $\sim$~1.3 MeV, corresponding to
the decay of the 1.78 MeV (${5/2}^+$) state in $^{11}$Be. There is
also another peak apparent near 2.2 MeV, corresponding to
decay of the 2.69 MeV  (3/2$^-$) state \cite{Millener2}. The very
narrow peak near threshold is compatible with decay from a state
at $\sim$4.0 MeV in $^{11}$Be to the first 2$^+$ state in $^{10}$Be at
3.37 MeV. The ground state branch of this decay corresponds to the peak at 
$E_{rel}\sim$3.5 MeV; its inclusion improves the fit, but the magnitude of
neither peak is well defined by the data for this $\sim$4.0 MeV state.
However, good candidates exist for such a state in $^{11}$Be 
\cite{BohlenMoscow,Ajzenberg1990}. 

The detection efficiency for neutrons from $^{11}$Be* decay is
determined in part by their laboratory angular distribution, which
in turn depends on the decay energy to $^{10}$Be + n and also the
spread in momentum induced by the initial neutron removal from $^{12}$Be.
Detailed simulations were performed \cite{thesis}, including the effects of the
geometric acceptance of the neutron detector array, the energy and angular
straggling of charged particles, the energy loss in the target,
and the divergence and energy spread of the beam. The effects of
the telescope resolution and efficiency were also
included, along with the absorption of neutrons by the telescope
(a 10\% effect \cite{Lecouey}). The momentum spread arising from the neutron
removal was determined from the measured angular distribution of
neutrons from the very low energy decay of $^{11}$Be*(4.0 MeV) to
$^{10}$Be*(2$^+$, 3.37 MeV) + n (cf. Figure \ref{fig-Erel}), where
the neutron momentum distribution was dominated by 
that of the $^{11}$Be* before decay. The detection of a
fast neutron ($E_n >$15 MeV) from the breakup of $^{12}$Be, measured in
coincidence with a $^{10}$Be from the subsequent decay of
$^{11}$Be* was also simulated. 

Simulations were performed for the decay of the unbound states in
$^{11}$Be below 4 MeV (isotropically in the $^{11}$Be rest frame), including the decay from the $\sim$4 MeV state
to the 2$^+$ state in $^{10}$Be, and also the detection of
neutrons diffracted from $^{12}$Be in coincidence with a $^{10}$Be
core. The simulated events were analyzed in the same manner as the
experimental data. The resulting $E_{rel}$ line-shapes (Fig. \ref{fig-Erel}) were
least-squares fitted to the experimental distribution and, using
the detection efficiency determined from the
simulations (Fig. \ref{fig-Erel}, upper panel) 
the cross sections for the different states
were determined. As a consistency check, the $^{11}$Be transverse momentum distribution (from $^{10}$Be+n), and the  neutron
angular distribution 
$\frac{d\sigma _n}{d\Omega}$ in coincidence
with $^{10}$Be were reconstructed from the simulated events, and in both cases the agreement with experiment was very good. 
The simulated ``hit'' efficiency for neutrons was checked and agreed with the value determined by fitting and integrating
$\frac{d\sigma _n}{d\Omega}$
to within 1\%.

The cross section extracted for the production of the 1.78 MeV
${5/2}^+$  state is included in Table \ref{tab-cross}, along with
that for the 2.69 MeV $3/2^-$ state and the bound $1/2^-$ state.
The estimated uncertainties include contributions from statistical fitting
plus uncertainties in the efficiency corrections, target
thickness and background subtraction. The cross section for
diffractive breakup to produce a bound $^{11}$Be (either 1/2$^+$
or 1/2$^-$) and a fast neutron was also extracted and was 46 $\pm$ 10 mb.

\begin{table*}
\caption{\label{tab-cross} Cross sections for states in $^{11}$Be
produced via neutron removal from $^{12}$Be on a carbon
target at 39.3 MeV/nucleon (present work) compared with 
reaction calculations and previous  work   at 78
MeV/nucleon on $^9$Be. Uncertainties for $S_{\rm exp}$ are experimental only (for comparison, ref.
\cite{Navin} values have been adjusted to remove the assumed 20\% theory uncertainty). }
\begin{ruledtabular}
\begin{tabular}{ccccccccccc}
$J^\pi$  & $E_{x}$  & $\sigma _{\rm exp}$ & $\sigma _{\rm strip}$& $\sigma _{\rm diff}$& $\sigma _{\rm sp}$
 & $S_{\rm exp}$ $^a$ & $S_{\rm exp}$ $^a$ & WBT2 & WBT2\,$^\prime $ & EXC2 \\
 &  (MeV) & (mb) & (mb) & (mb) &  (mb) & (present work)  & Ref. \cite{Navin} & Ref. \cite{Navin} & (see text) & Ref. \cite{Nunes}\\
\hline
$1/2^+$ & 0.00 & $^b$ & 61.28 & 57.69 & 118.96 & [0.56 $\pm$ 0.18]$^b$ & 0.42 $\pm$ 0.05 & 0.69 & 0.55 & 0.44\\
$1/2^-$ & 0.32 & 32.5 $\pm$ 6.1 & 42.59 & 30.72 & 73.31 & 0.44 $\pm$ 0.08 & 0.37 $\pm$ 0.07 & 0.58 & 0.47 & 0.38\\
$5/2^+$ & 1.78 & 30.3 $\pm$ 4.0 & 39.78 & 22.77 & 62.55 & 0.48 $\pm$ 0.06 & - & 0.55 & 0.44 & 0.58\\
$3/2^-$ & 2.69 & 22.6 $\pm$ 4.1 &  35.47 & 20.87 & 56.35 & 0.40 $\pm$ 0.06 & - & & \\
\end{tabular}
\end{ruledtabular}
$^a$uncertainties do not include contribution from theoretical model of reaction mechanism, estimated to be $\pm$10-20\%. \hfill \mbox{} \\
$^b$total $\sigma _{\rm exp}$ not measured, but $\sigma _{\rm diff} = 46 \pm 10$ mb for 1/2$^+$ and 1/2$^-$ together and $S_{\rm exp}$ deduced from this (see text). \hfill \mbox{}
\end{table*}

The measured neutron removal partial cross sections from $^{12}$Be
can be interpreted in terms of spectroscopic factors using a
reaction model. The spectroscopic factors listed in Table
\ref{tab-cross} are, except for the 1/2$^+$ state, the ratio of
the experimental ($\sigma _{\rm exp}$) to the eikonal model partial cross section ($\sigma _{\rm sp}$). For
the 1/2$^+$ state, the measurement of diffractive breakup of
$^{12}$Be to the two bound $^{11}$Be states has been used, with the 1/2$^-$
contribution being subtracted according to its theoretical
diffraction cross section (see Table \ref{tab-cross}) and the spectroscopic factor 
of $0.44 \pm 0.08$ found here. Note that the uncertainties quoted for $S_{\rm exp}$ in Table \ref{tab-cross} 
do not include any contribution arising from the assumptions in the reaction calculation, where for example
two-step processes are not included, and it is estimated that this implies an additional uncertainty
of up to 20\% (which is consistent with ref. \cite{Navin}).

The present reaction analysis follows closely the eikonal model of Refs.\
\cite{Navin,two}. The neutron-target S-matrix was
computed from the target density and the 
JLM effective nucleon-nucleon interaction \cite{three}. The
usual real and imaginary part scale factors ($\lambda_V =1.0$,
$\lambda_W = 0.8$) were applied to the optical potential. The
matter densities for $^{12}$C and $^{10}$Be were of harmonic
oscillator and Gaussian form, respectively, with rms radii of 2.4
fm and 2.28 fm. For the $1/2^+$, $1/2^-$ and $5/2^+$ (particle)
transitions, ($^{10}$Be+n) composite core-target S-matrices were
constructed from those of $^{10}$Be and the neutron \cite{two}. In
the (unbound) $5/2^+$ case, a separation energy of 0.01 MeV was
used to compute the S-matrix. For the $3/2^-$ (hole) state, a
Gaussian (mass 11) core density of rms radius 2.54 fm was used,
representative of the size of $^{12}$Be. The removed-nucleon
single-particle overlaps were taken as eigenstates of Woods-Saxon
potentials, with geometry $r_0$=1.25 fm and $a$=0.7 fm, and with
depths adjusted to the physical separation energies for each final
state.

Our observation of the $5/2^+$ state in knockout from $^{12}$Be is the first direct experimental evidence for a significant $d$-wave intruder component in any $N$=8 isotone. Barker \cite{Barker1}
first pointed out that the observed lowering of the $1s_{1/2}$ and
$0d_{5/2}$ neutron orbitals in $^{11}$Be should lead to
strong admixtures of both these orbitals in the ground state of
$^{12}$Be and
concluded \cite{Barker1, Barker2} that as little as 20-40\%
of the $^{12}$Be ground state might comprise the $0\hbar \omega$
$\nu(0p_{1/2})^2$ configuration. For $^{11}$Li, this model also
successfully predicted a strong $1s_{1/2}$ strength admixture in the ground state.

The $(1s0d)^2$ contribution to the $^{12}$Be ground state was deduced
in an indirect fashion by Fortune {\em et~al.} \cite{Fortune} in the light of
their measurements of $\tau _{1/2}$ and E($2_1^+$) for $^{12}$Be
\cite{Alburger1,Alburger2}. Indeed, according to subsequent
\emph{psd}-space shell model calculations, the $\beta$-decay from
the $^{12}$Be ground state is quenched to an extent that places an
upper limit of 35\% on the $0\hbar \omega$ component of the
wavefunction   \cite{Suzuki1}. The $^{12}$O-$^{12}$Be Coulomb
energy difference also supports shell breaking of this order
\cite{SherrFortune1}.

The magnitude of shell breaking observed in the present work can
be quantified by comparing with theory. Table \ref{tab-cross}
includes spectroscopic factors (WBT2) based on the shell
model calculations reported in ref. \cite{Navin}, which
correspond to a mixing of 32\% $0\hbar \omega$ ($0p^8$) and 68\%
$2\hbar \omega$ ($0p^6(1s0d)^2$) contributions. 
The sum of these is close to the value 2.0 that would be expected in the simplest picture.
These values are scaled by 0.8 to give WBT2\,$^\prime$ (which is consistent with other knockout work \cite{Hansen}) and this reproduces the present experimental results very well. 
Also listed for comparison, the EXC2 values from a 3-body model including $^{10}$Be core excitation \cite{Nunes} give good agreement with the present work while reproducing other features of $^{12}$Be.

Recently, Iwasaki \emph{et al.} inferred the
disappearance of the $N$=8 shell gap in $^{12}$Be from both the
deformation length derived from inelastic proton scattering to the
$2_1^+$ state \cite{Iwasaki1}, and the low energy and large
$B$(E1;$0^+ \rightarrow 1^-$) for the $1_1^-$ state
\cite{Iwasaki2}.
Their deductions agreed in detail with the \emph{psd} shell model
calculations \cite{Suzuki1,Sagawa}, concluding that the $0p_{1/2}$ and
$1s_{1/2}$ orbitals are effectively degenerate for $^{12}$Be, just
as they are in $^{11}$Be.

Consistent theoretical results are obtained using nuclear field
theory \cite{Gori}, which predicts both the presence of large
$\nu(0d_{5/2})^2$ strength in $^{12}$Be and its absence in
$^{11}$Li, in agreement with Barker \cite{Barker2} and with a recent theoretical analysis
\cite{Bertulani} of $^{11}$Li reaction data \cite{HaikSimon}. It is
interesting to note that the early work of Barker
\cite{Barker1} also predicted a low-lying $0_2^+$ state at 2.35
MeV in $^{12}$Be which has only recently been observed, at 2.24 MeV \cite{Shimoura1}.



Thus, the neutron shell breaking observed at $N$=8 for $^{12}$Be, but not
$^{14}$C, is similar to the breaking of $N$=20 for $^{32}$Mg but not
$^{34}$Si. In each case, when the proton $j_{>}$ orbital is full,
the neutrons are magic. When protons are removed, the full $j_{<}$
orbital for neutrons is no longer magic, the next shell
intrudes and deformation results.

In conclusion, one-neutron removal cross sections from $^{12}$Be
to the 0.32 MeV ($1/2^-$) and 1.78 MeV ($5/2^+$) states in
$^{11}$Be have been measured. From these and eikonal model
calculations, spectroscopic factors were deduced. These indicate
strong breaking of the $N$=8 magic number in $^{12}$Be, including a
significant $d$-wave component. This is distinct from other $N$=8
isotones and is the first direct experimental confirmation of the
predictions of a number of structure models.

The authors wish to acknowledge the support provided by the technical staff of LPC and GANIL.
Partial support through the EU Human Mobility programme of the European Community is also acknowledged.

\bibliography{be12-310805}

\begin{thebibliography}{32}
\expandafter\ifx\csname natexlab\endcsname\relax\def\natexlab#1{#1}\fi
\expandafter\ifx\csname bibnamefont\endcsname\relax
  \def\bibnamefont#1{#1}\fi
\expandafter\ifx\csname bibfnamefont\endcsname\relax
  \def\bibfnamefont#1{#1}\fi
\expandafter\ifx\csname citenamefont\endcsname\relax
  \def\citenamefont#1{#1}\fi
\expandafter\ifx\csname url\endcsname\relax
  \def\url#1{\texttt{#1}}\fi
\expandafter\ifx\csname urlprefix\endcsname\relax\def\urlprefix{URL }\fi
\providecommand{\bibinfo}[2]{#2}
\providecommand{\eprint}[2][]{\url{#2}}

\bibitem[{\citenamefont{Simon et~al.}(1999)}]{HaikSimon}
\bibinfo{author}{\bibfnamefont{H.}~\bibnamefont{Simon}} \bibnamefont{et~al.},
  \bibinfo{journal}{Phys. Rev. Lett.} \textbf{\bibinfo{volume}{83}},
  \bibinfo{pages}{496} (\bibinfo{year}{1999}).

\bibitem[{\citenamefont{Iwasaki et~al.}(2000{\natexlab{a}})}]{Iwasaki1}
\bibinfo{author}{\bibfnamefont{H.}~\bibnamefont{Iwasaki}} \bibnamefont{et~al.},
  \bibinfo{journal}{Phys. Lett. B} \textbf{\bibinfo{volume}{481}},
  \bibinfo{pages}{7} (\bibinfo{year}{2000}{\natexlab{a}}).

\bibitem[{\citenamefont{Navin et~al.}(2000)}]{Navin}
\bibinfo{author}{\bibfnamefont{A.}~\bibnamefont{Navin}} \bibnamefont{et~al.},
  \bibinfo{journal}{Phys. Rev. Lett.} \textbf{\bibinfo{volume}{85}},
  \bibinfo{pages}{266} (\bibinfo{year}{2000}).

\bibitem[{\citenamefont{Iwasaki et~al.}(2000{\natexlab{b}})}]{Iwasaki2}
\bibinfo{author}{\bibfnamefont{H.}~\bibnamefont{Iwasaki}} \bibnamefont{et~al.},
  \bibinfo{journal}{Phys. Lett. B} \textbf{\bibinfo{volume}{491}},
  \bibinfo{pages}{8} (\bibinfo{year}{2000}{\natexlab{b}}).

\bibitem[{\citenamefont{Shimoura et~al.}(2003)}]{Shimoura1}
\bibinfo{author}{\bibfnamefont{S.}~\bibnamefont{Shimoura}}
  \bibnamefont{et~al.}, \bibinfo{journal}{Phys. Lett. B}
  \textbf{\bibinfo{volume}{560}}, \bibinfo{pages}{31} (\bibinfo{year}{2003}).

\bibitem[{\citenamefont{{F.C. Barker}}(1976)}]{Barker1}
\bibinfo{author}{\bibnamefont{{F.C. Barker}}}, \bibinfo{journal}{J. Phys. G}
  \textbf{\bibinfo{volume}{2}}, \bibinfo{pages}{L45} (\bibinfo{year}{1976}).

\bibitem[{\citenamefont{{F.C. Barker} and {G.T. Hickey}}(1977)}]{Barker2}
\bibinfo{author}{\bibnamefont{{F.C. Barker}}} \bibnamefont{and}
  \bibinfo{author}{\bibnamefont{{G.T. Hickey}}}, \bibinfo{journal}{J. Phys. G}
  \textbf{\bibinfo{volume}{3}}, \bibinfo{pages}{L23} (\bibinfo{year}{1977}).

\bibitem[{\citenamefont{{D.E. Alburger}
  et~al.}(1978{\natexlab{a}})}]{Alburger2}
\bibinfo{author}{\bibnamefont{{D.E. Alburger}}} \bibnamefont{et~al.},
  \bibinfo{journal}{Phys. Rev. C} \textbf{\bibinfo{volume}{18}},
  \bibinfo{pages}{2727} (\bibinfo{year}{1978}{\natexlab{a}}).

\bibitem[{\citenamefont{{H.T. Fortune} et~al.}(1994)}]{Fortune}
\bibinfo{author}{\bibnamefont{{H.T. Fortune}}} \bibnamefont{et~al.},
  \bibinfo{journal}{Phys. Rev. C} \textbf{\bibinfo{volume}{50}},
  \bibinfo{pages}{1355} (\bibinfo{year}{1994}).

\bibitem[{\citenamefont{{B.A. Brown}}(2001)}]{BABProgReview}
\bibinfo{author}{\bibnamefont{{B.A. Brown}}}, \bibinfo{journal}{Prog. Part.
  Nucl. Phys.} \textbf{\bibinfo{volume}{47}}, \bibinfo{pages}{517}
  (\bibinfo{year}{2001}).

\bibitem[{\citenamefont{{Y. Kanada-En'yo} and {H. Horiuchi}}(2003)}]{Kanada2}
\bibinfo{author}{\bibnamefont{{Y. Kanada-En'yo}}} \bibnamefont{and}
  \bibinfo{author}{\bibnamefont{{H. Horiuchi}}}, \bibinfo{journal}{Phys. Rev.
  C} \textbf{\bibinfo{volume}{68}}, \bibinfo{pages}{014319}
  (\bibinfo{year}{2003}).

\bibitem[{\citenamefont{Gori et~al.}(2004)}]{Gori}
\bibinfo{author}{\bibfnamefont{G.}~\bibnamefont{Gori}} \bibnamefont{et~al.},
  \bibinfo{journal}{Phys. Rev. C} \textbf{\bibinfo{volume}{69}},
  \bibinfo{pages}{2004} (\bibinfo{year}{2004}).

\bibitem[{\citenamefont{{H. Sagawa} et~al.}(2001)}]{Sagawa}
\bibinfo{author}{\bibnamefont{{H. Sagawa}}} \bibnamefont{et~al.},
  \bibinfo{journal}{Phys. Rev. C} \textbf{\bibinfo{volume}{63}},
  \bibinfo{pages}{034310} (\bibinfo{year}{2001}).

\bibitem[{\citenamefont{{F. Nunes} et~al.}(2002)}]{Nunes}
\bibinfo{author}{\bibnamefont{{F. Nunes}}} \bibnamefont{et~al.},
  \bibinfo{journal}{Nucl. Phys. A} \textbf{\bibinfo{volume}{703}},
  \bibinfo{pages}{593} (\bibinfo{year}{2002}).

\bibitem[{\citenamefont{{J.C. Pachero} and {N. Vinh Mau}}(2002)}]{VinhMau}
\bibinfo{author}{\bibnamefont{{J.C. Pachero}}} \bibnamefont{and}
  \bibinfo{author}{\bibnamefont{{N. Vinh Mau}}}, \bibinfo{journal}{Phys. Rev.
  C} \textbf{\bibinfo{volume}{65}}, \bibinfo{pages}{044004}
  (\bibinfo{year}{2002}).

\bibitem[{\citenamefont{{B. Zwieglinski} et~al.}(1979)}]{Zwieglinski}
\bibinfo{author}{\bibnamefont{{B. Zwieglinski}}} \bibnamefont{et~al.},
  \bibinfo{journal}{Nucl. Phys. A} \textbf{\bibinfo{volume}{315}},
  \bibinfo{pages}{124} (\bibinfo{year}{1979}).

\bibitem[{\citenamefont{{N. Fukuda} et~al.}(2004)}]{Fukuda1}
\bibinfo{author}{\bibnamefont{{N. Fukuda}}} \bibnamefont{et~al.},
  \bibinfo{journal}{Phys. Rev. C} \textbf{\bibinfo{volume}{70}},
  \bibinfo{pages}{054606} (\bibinfo{year}{2004}).

\bibitem[{\citenamefont{{F. Ajzenberg-Selove}}(1990)}]{Ajzenberg1990}
\bibinfo{author}{\bibnamefont{{F. Ajzenberg-Selove}}}, \bibinfo{journal}{Nucl.
  Phys. A} \textbf{\bibinfo{volume}{506}}, \bibinfo{pages}{1}
  (\bibinfo{year}{1990}).

\bibitem[{\citenamefont{{S. Ahmed} et~al.}(2004)}]{Ahmed}
\bibinfo{author}{\bibnamefont{{S. Ahmed}}} \bibnamefont{et~al.},
  \bibinfo{journal}{Phys. Rev. C} \textbf{\bibinfo{volume}{69}},
  \bibinfo{pages}{024303} (\bibinfo{year}{2004}).

\bibitem[{\citenamefont{Tilquin et~al.}(1995)}]{Tilquin}
\bibinfo{author}{\bibfnamefont{I.}~\bibnamefont{Tilquin}} \bibnamefont{et~al.},
  \bibinfo{journal}{Nucl. Inst. Meth. A} \textbf{\bibinfo{volume}{365}},
  \bibinfo{pages}{446} (\bibinfo{year}{1995}).

\bibitem[{\citenamefont{{M. Labiche} et~al.}(2001)}]{LabichePRL}
\bibinfo{author}{\bibnamefont{{M. Labiche}}} \bibnamefont{et~al.},
  \bibinfo{journal}{Phys. Rev. Lett.} \textbf{\bibinfo{volume}{86}},
  \bibinfo{pages}{600} (\bibinfo{year}{2001}).

\bibitem[{\citenamefont{{S.D. Pain}}(2004)}]{thesis}
\bibinfo{author}{\bibnamefont{{S.D. Pain}}}, Ph.D. thesis,
  \bibinfo{school}{University of Surrey} (\bibinfo{year}{2004}).

\bibitem[{\citenamefont{{D.J. Millener}}(2001)}]{Millener2}
\bibinfo{author}{\bibnamefont{{D.J. Millener}}}, \bibinfo{journal}{Nucl. Phys.
  A} \textbf{\bibinfo{volume}{693}}, \bibinfo{pages}{394}
  (\bibinfo{year}{2001}).

\bibitem[{\citenamefont{{H.G. Bohlen} et~al.}(2004)}]{BohlenMoscow}
\bibinfo{author}{\bibnamefont{{H.G. Bohlen}}} \bibnamefont{et~al.},
  \bibinfo{journal}{Nucl. Phys. A} \textbf{\bibinfo{volume}{734}},
  \bibinfo{pages}{345} (\bibinfo{year}{2004}).

\bibitem[{\citenamefont{{J.L. Lecouey}}(2003)}]{Lecouey}
\bibinfo{author}{\bibnamefont{{J.L. Lecouey}}}, Ph.D. thesis,
  \bibinfo{school}{Universit\'e de Caen} (\bibinfo{year}{2003}).

\bibitem[{\citenamefont{{J.A. Tostevin}}(2001)}]{two}
\bibinfo{author}{\bibnamefont{{J.A. Tostevin}}}, \bibinfo{journal}{Nucl. Phys.
  A} \textbf{\bibinfo{volume}{682}}, \bibinfo{pages}{320c}
  (\bibinfo{year}{2001}).

\bibitem[{\citenamefont{Jeukenne et~al.}(1977)\citenamefont{Jeukenne, Lejeune,
  and Mahaux}}]{three}
\bibinfo{author}{\bibfnamefont{J.}~\bibnamefont{Jeukenne}},
  \bibinfo{author}{\bibfnamefont{A.}~\bibnamefont{Lejeune}}, \bibnamefont{and}
  \bibinfo{author}{\bibfnamefont{C.}~\bibnamefont{Mahaux}},
  \bibinfo{journal}{Phys. Rev. C} \textbf{\bibinfo{volume}{16}},
  \bibinfo{pages}{80} (\bibinfo{year}{1977}).

\bibitem[{\citenamefont{{D.E. Alburger}
  et~al.}(1978{\natexlab{b}})}]{Alburger1}
\bibinfo{author}{\bibnamefont{{D.E. Alburger}}} \bibnamefont{et~al.},
  \bibinfo{journal}{Phys. Rev. C} \textbf{\bibinfo{volume}{17}},
  \bibinfo{pages}{1523} (\bibinfo{year}{1978}{\natexlab{b}}).

\bibitem[{\citenamefont{Suzuki and Otsuka}(1997)}]{Suzuki1}
\bibinfo{author}{\bibfnamefont{T.}~\bibnamefont{Suzuki}} \bibnamefont{and}
  \bibinfo{author}{\bibfnamefont{T.}~\bibnamefont{Otsuka}},
  \bibinfo{journal}{Phys. Rev. C} \textbf{\bibinfo{volume}{56}},
  \bibinfo{pages}{847} (\bibinfo{year}{1997}).

\bibitem[{\citenamefont{Sherr and {H.T. Fortune}}(1999)}]{SherrFortune1}
\bibinfo{author}{\bibfnamefont{R.}~\bibnamefont{Sherr}} \bibnamefont{and}
  \bibinfo{author}{\bibnamefont{{H.T. Fortune}}}, \bibinfo{journal}{Phys. Rev.
  C} \textbf{\bibinfo{volume}{60}}, \bibinfo{pages}{64323}
  (\bibinfo{year}{1999}).

\bibitem[{\citenamefont{{P.G. Hansen and J.A. Tostevin}}(2003)}]{Hansen}
\bibinfo{author}{\bibnamefont{{P.G. Hansen and J.A. Tostevin}}},
  \bibinfo{journal}{Annu. Rev. Nucl. Part. Sci.} \textbf{\bibinfo{volume}{53}},
  \bibinfo{pages}{219} (\bibinfo{year}{2003}).

\bibitem[{\citenamefont{{C.A. Bertulani} and {P.G. Hansen}}(2004)}]{Bertulani}
\bibinfo{author}{\bibnamefont{{C.A. Bertulani}}} \bibnamefont{and}
  \bibinfo{author}{\bibnamefont{{P.G. Hansen}}}, \bibinfo{journal}{Phys. Rev.
  C} \textbf{\bibinfo{volume}{70}}, \bibinfo{pages}{0343609}
  (\bibinfo{year}{2004}).

\end{thebibliography}

\end{document}